# Software paper for submission to the Journal of Open Research Software

To complete this template, please replace the blue text with your own. The paper has three main sections: (1) Overview; (2) Availability; (3) Reuse potential.

Please submit the completed paper to: editor.jors@ubiquitypress.com

---

## (1) Overview

### Title
ExpressInHost: A codon tuning tool for the expression of recombinant proteins in host microorganisms

### Paper Authors
1. Raguin, Adélaïde;
2. Stansfield, Ian;
3. Romano, Maria Carmen.

### Paper Author Roles and Affiliations
1. Adélaïde Raguin: conceptualisation and formulation, programming, development and testing of the software, data collection, writing and reviewing of the manuscript.
Department of Computer Sciences, Institute of Computational Cell Biology, Heinrich-Heine University, 40225 Düsseldorf, Germany

2. Ian Stansfield: conceptualisation and formulation of the ideas, data collection, supervision, project administration and funding acquisition.
Institute of Medical Sciences, School of Medicine, Medical Sciences and Nutrition, University of Aberdeen, Aberbeen, AB25 2ZD, UK

3. Maria Carmen Romano: conceptualisation and formulation of the ideas, reviewing of the manuscript, supervision, project administration and funding acquisition.
Institute for Complex Systems and Mathematical Biology, University of Aberdeen, Aberdeen, AB24 3UE, UK
Institute of Medical Sciences, School of Medicine, Medical Sciences and Nutrition, University of Aberdeen, Aberbeen, AB25 2ZD, UK

### Abstract
ExpressInHost (https://gitlab.com/a.raguin/expressinhost) is a GTK/C++ based user friendly graphical interface that allows tuning the codon sequence of an mRNA for recombinant protein expression in a host microorganism. Heterologous gene expression is widely implemented in biotechnology companies and academic research laboratories. However, expression of recombinant proteins can be challenging. On the one hand, maximising translation speed is important, especially in scalable production processes relevant to biotechnology companies, but on the other hand, solubility problems often arise as a consequence, since translation 'pauses' might be key to allow the nascent polypeptide chain to fold appropriately. To address this challenge, we have developed a software that offers three distinct modes to tune codon sequences using the genetic code redundancy. The tuning strategies implemented take into account the specific tRNA resources of the host and that of the native organism. They balance rapid translation and native speed mimicking, which might be important to allow proper protein folding, thereby avoiding protein solubility problems.

**Introduction**

Recombinant protein expression consists in taking a gene of interest (called target gene) from an organism (called native organism) and to express it in another organism (called host organism). The purpose of recombinant protein expression is to extract and purify quickly large amounts of a target protein expressed in a selected host, for instance, an engineered microorganism. Recombinant proteins are complex and large molecules in comparison to traditional chemically produced drugs, allowing more sophisticated biochemical activity [1]. These properties open the door to new and successful therapy strategies, which started in 1982 with the first production of human insulin in *Escherichia coli* by Genentech [2]. In addition to pharmacology, recombinant proteins are widely used in industrial applications, from food additive [3], to glue and biofuel production [4]. Therefore, they are nowadays central in the development of biotechnologies. Although in theory recombinant protein expression is a straightforward process, in practice many pitfalls can surface [5]. To tackle them, tools derived from bioinformatics represent an attractive alternative that complements experimental studies.

Several software tools are available online for codon tuning. A number of them are patented and directly serve the commercial purpose of selling the tuned gene plasmids (e.g. GenSmart Design [6] and Genewiz [7]). In contrast, some other software tools are freely and openly released and sometimes accessible from online platforms [8] or downloadable as packages [9]. In between, one can find algorithms that are not open-source [10,11], or should be requested from the authors [12]. For a more complete view on available software tools to optimise protein expression, please see [13,14]. Many of those protein expression optimisation software tools are based on the tRNA Adaptation Index (tAI) [15]. It relates a codon to the abundance of the cognate tRNA, and it was inspired by the work of Sharp and Li on the Codon Adaptation Index (CAI) [16].

The latter determines whether a codon is present in highly expressed genes, and is motivated by seminal work initiated in the early 1980s, by pioneers like Ikemura, [17-19] Gouy and Gautier [20], and Bennetzen and Hall [21]. Here we follow a parallel approach, where we propose a software tool, ExpressInHost, for recombinant protein synthesis optimisation based on two dimensionless quantities, namely the translation "speed index" and translation "rank index". Both indices are based on tRNA Gene Copy Numbers and they allow us to analyse gene translation profiles across organisms. Specifically, with ExpressInHost, we do not only consider the native organism, but also its host, as well as other organisms. The central focus of ExpressInHost is to increase the recombinant protein synthesis rate while preserving proper protein folding. Towards it, we propose various hypotheses, that correspond to the distinct "Modes" of the software tool. Our overarching aim is to release as widely as possible this tool, such that our hypotheses can be tested in experimental set ups, for validation, and potential further developments.

ExpressInHost is a fully open source and freely accessible software that has been developed to address the challenge of recombinant protein expression. The software tool has been written with the programming language GTK/C++, and it is fully designed in a user-friendly manner. The graphical user interface is complemented with an "Instructions window" that guides the user step by step. This tool results from our research project on predictive optimisation of biocatalyst production for high-value chemical manufacturing. The project focussed on heterologous protein production optimisation in *Escherichia coli*. Our cross-platform software can be further developed, and above all, it can easily be utilised by both industrial and academic groups without any programming background being required.

## Model and underlying hypotheses

### Translation speed profiles

The speed at which ribosomes translate is non-uniform along mRNAs, and several factors have been identified to influence this complex and not fully understood process. For instance, it has been shown that mRNA secondary structure is generally not a central factor [22]. Instead, the abundance of charged tRNAs is known to influence the ribosome decoding speed at each codon [23]. As a proxy, we assume that fast codons are decoded by abundant tRNAs while slow codons are decoded by rare tRNAs.

The model of translation that supports ExpressInHost [24-26] assumes that the time a ribosome needs to decode a certain codon is on average proportional to the abundance of its cognate tRNA. Since different tRNAs occur typically in different abundances in the cytoplasm, codons are assigned a translation speed index which depends on the concentration of their cognate tRNAs. It has been shown that the abundance of a certain tRNA correlates with its Gene

Copy Number (GCN) [27]. Hence, the GCN is commonly used as a proxy for tRNA concentration.

Experimental data suggest that the translation rates of codons using the G-U wobble are reduced by 39% compared to their G-C counterparts. Analogously, codons using the wobble I-C and codons using the wobble I-A are reduced by 36% relative to their I-U counterparts [28,29]. For simplicity, in ExpressInHost a unique reduction rate applies for the calculation of the wobble base-pairing. The translation speed index of a wobbly codon is reduced by 35% as compared to that of the counterpart cognate codon.

In ExpressInHost, translation speed index is therefore assumed to be proportional to the GCN of the corresponding tRNA, also considering wobble base-pairing. We assign a translation speed index $s_i$ to each codon as follows:

$$s_i = \frac{GCN_i * W_i}{\sum_{j=0}^{T} GCN_j * W_j}, \qquad (1)$$

where $GCN_i$ is the GCN of the tRNA decoding the codon i, and $W_i$ takes the value 0,65 if the tRNA binds to the codon using wobble base-pairing, and 1, otherwise. The sum in the denominator is taken over all 61 codons so that the translation speed indices assigned to each codon are in the range [0;1].

Additionally, we define a rank index $r_i$ to rank the speed indices of synonymous codons for a specific amino acid. Some synonymous codons are translated by different tRNAs (e.g. in *S. cerevisiae*, there are 41 different tRNAs and 20 amino acids). For each amino acid, the rank index is normalised between 0 (slowest synonymous codon) and 1 (fastest synonymous codon), such that the rank index of any codon can be compared in different organisms. The rank index $r_i$ of a codon is defined as:

$$r_i = \frac{GCN_i * W_i - GCN_{min} * W_{min}}{GCN_{max} * W_{max} - GCN_{min} * W_{min}}, \qquad (2)$$

where i runs from 1 to the number N of synonymous codons coding for the amino acid under consideration. $GCN_{min}$ and $GCN_{max}$ are the minimum and maximum GCN, respectively, of the isoacceptor tRNAs (decoding the same amino acid).

The purpose of the translation speed index $s_i$ is to evaluate how slow (or fast) is a codon within a certain organism, while the rank index $r_i$ allows comparing exclusively synonymous codons, and within different organisms.

**Protein folding**
The impact of slow and fast codons on protein folding has been reviewed in [23]. In particular it has been found that: i) substitutions with synonymous codons that invert the programmed speed of mRNA translation, from either fast to slow or vice versa, are deleterious for folding and expression of the encoded protein [30-32]; ii) the distribution of slow-translated codons is precisely selected at specific positions along mRNA to facilitate co-translational folding and translocation of the encoded protein [33-36]; and iii) regions of slow or fast translation in homologous proteins are conserved among species [37], whereby the selection of the nucleotide sequence is driven by preserving the translation pattern rather than conservation of the codon or amino acid identities [38].

Correct co-translational folding is key to guarantee proper protein function and avoid solubility problems. However, translation pausing does not necessarily correlate with correct folding, and long pausing can instead lead to translation mistakes, like frameshifting [39]. Therefore, the usage of slow codons and the patterning of translation speed profiles is a highly complex and not fully understood challenge. A simple working hypothesis is that it should satisfy apparently contradictory constraints: translation speed should be high to maximise production, but locally, it should be slow enough to allow for proper protein folding, without being too slow to avoid typical translation mistakes. In ExpressInHost, we assume that it is thus essential to control both the positioning and the strength of the pausing, in order to avoid mistranslation and misfolding.

**Tuning translation speed for recombinant protein synthesis**
As discussed above, the translation speed profile of an mRNA depends on the abundance of the tRNAs, and it is suggested to impact on protein folding. As the abundance of different tRNAs varies with the organism under consideration, in general, the translation speed profile of a certain mRNA will be different in a host organism compared to the one in the native organism. Considering the hypotheses taken for ExpressInHost, if we were only interested in maximising translation speed, we could simply choose the fastest synonymous codon for each amino acid along the mRNA, therefore taking into consideration only the tRNA abundances of the host organism. However, as mentioned above, that would potentially miss key translation positions that aid proper protein folding.

Importantly, in addition to the "pauses" (slow translation speed positions) reported in the experimental literature, in ExpressInHost we also formulate other hypotheses to identify key positions that might aid proper co-translational protein folding. The tuning procedures examine the translation speed profile of the mRNA in the native organism, and consider orthologous genes across a number of different organisms, to perform a similarity analysis among them. Where key positions are identified, instead of enforcing slowest possible speed, that could for instance lead to translation mistakes as reported above, we choose to mimic the translation speed as it has been evolved in the native organism. Therefore, in order to optimise heterologous protein expression, while at the same time preserving key translation positions that aid proper co-translational folding, we postulate that it is not enough to consider the abundances of tRNAs in the host organism, as typically done in other standard tools. To address this matter and propose alternative hypotheses, we developed three different codon tuning strategies for recombinant protein synthesis.

**Mode 1: Direct mapping.** This tuning mode mimics the translation speed profile from the native organism into the host organism, for all codons along the mRNA. The assumption underlying this mode is that the native organism has naturally achieved the optimal trade-off between speed and accurate folding. Hence, we map the native translation speed profile into the host by choosing synonymous codons that have the closest possible rank index to the native ones.

**Mode 2: Optimisation and conservation I.** This tuning mode considers a protein sequence similarity analysis to identify conserved amino acids across a set of orthologous proteins from different organisms. The underlying assumption is that those conserved amino acids play a crucial role in protein function, and therefore, it is especially important to retain the native translation speed of their corresponding codons into the host organism. We call such a position along the sequence "conserved amino acid", and we mimic the speed of the native codon in the tuned sequence by choosing a synonymous codon that has the closest possible rank index to the native one. For the rest of the codons along the sequence, this mode maximises the speed, i.e. it chooses the fastest possible synonymous codon (rank index = 1).

**Mode 3: Optimisation and conservation II.** This tuning mode also considers a set of orthologous proteins, but it identifies key co-translational folding positions along the mRNA sequence in a different way. It individually analyses the translation speed profile of each sequence in the set of orthologous proteins, and it determines where a slow translation codon (low speed index) is consistently used across the different organisms. For such a position, this tuning mode retains the native translation speed in the tuned sequence by choosing the synonymous codon that has the closest possible rank index to the native one. For the rest of the codons along the sequence, this mode maximises the speed, i.e. it chooses the fastest possible synonymous codon (rank index = 1). This mode therefore assumes that certain slow codons are crucial for protein folding, and that those are conserved throughout orthologous genes expressed in different organisms.

It should be noted that the different modes of ExpressInHost "tune" the nucleotide sequences provided by the user, but do not necessarily "optimise" them. For clarity, throughout this paper we call "optimised codons" exclusively those whose translation speed is maximised by the software (rank index = 1). Additionally, we call native sequences and native organisms all sequences and organisms that are input by the user, and we call tuned sequences those that are output by the software. The host organism is selected by the user as the microorganism in which input sequences are to be tuned.

**Implementation and architecture**

A unique variant of the software is released. It is released as an executable file for Windows users (with all dependencies for the specific libraries of the graphical interface) and as source codes for Linux users. The software is deposited on Gitlab and Zenodo with a detailed `Readme.md` file, together with example files.

The software is structured in three distinct `cpp` codes that have clear distinct roles as illustrated in Figure 1.

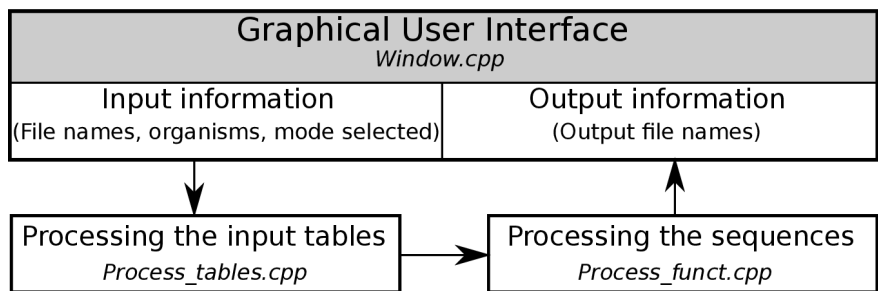

Figure 1: Diagram of the architecture of ExpressInHost.

The graphical user interface (GUI) (see Figure 2) is meant for the user to specify the input information, select the mode of the software, and collect output information. The input information requested depends on the tuning mode to be selected. The use of the GUI is explained in full detail in the expandable "Instructions window" (see Figure 3) that pops up as the software is launched. In brief, steps 1 to 5 of the GUI are used to input data on: the aligned amino acid sequences (relevant for tuning modes 2 and 3) (step 1), all nucleotide sequences (i.e. the target mRNA, and the orthologous ones used for the similarity analysis (relevant for tuning modes 2 and 3)) (step 2), the list of all native organisms (that of the target mRNA, and those of the orthologous proteins (relevant for tuning modes 2 and 3)) (steps 3 and 4), and the host organism (5). Steps 6 and 7 are used to select the tuning mode and to trigger the tuning of the sequences, respectively. Steps 8 and 9 are used to display the outcomes of the tuning process.

The input tables needed for the tuning process are tables containing the GCN of the different tRNAs for each of the organisms considered, i.e. the host, and all native organisms (that of the target mRNA, and those of the orthologous proteins (relevant for tuning modes 2 and 3)). The code `Process_tables.cpp` uses those tables to compute the translation speed index $s_i$ of each codon (Eq. 1) and its rank index $r_i$ (Eq. 2) for each organism. Additionally, for each codon, `Process_tables.cpp` determines whether it is slow or not in each organism in the following way: i) the median translation speed index is calculated for the organism, ii) the speed index range between the slowest codon and the median is calculated, iii) the threshold is set as 50% of that range, and iv) a codon is designated as slow if its speed index lies between the slowest codon and the threshold.

A library of those tRNA data tables is provided for several native organisms, and it is accessed by the user in a dropdown menu in step 3. The GUI also includes the possibility for the user to call non-listed native organisms in step 4. To do so, the user needs to construct the input table of tRNA data, in which codons decoded by wobble base-pairing should be indicated [28,29]. The construction of input tRNA tables for alternative organisms is supported by template tables, and it is fully guided in the `Readme.md` file accompanying the software.

`Process_funct.cpp` performs the tuning of the target mRNA sequence and the orthologous ones according to the mode selected by the user. The main assumptions underpinning each of the three tuning modes are detailed in the Introduction. Here, we describe the main algorithmic steps followed in each of these modes.

**Mode 1: Direct mapping.** In this tuning mode, `Process_funct.cpp`: i) reads the target sequence codon by codon; ii) reads the rank index of each codon calculated by `Process_tables.cpp` from the tRNA table of the native organism; and iii) selects the synonymous codon with the closest rank index calculated `Process_tables.cpp` from the tRNA table of the host organism.

**Mode 2: Optimisation and conservation I.** In this tuning mode, `Process_funct.cpp` requests the user to provide an input file that contains the target mRNA sequence, and a set of orthologous sequences. In addition to this set of nucleotide sequences, the user must provide the corresponding aligned protein sequences, e.g. using Clustal [40], a freely available online tool developed by the European Bioinformatics Institute (EBML-BBI) that aligns sequences based on seeded guide trees (see Figure 4). In addition to aligning the sequences, Clustal indicates the degree of conservation of amino acids at every position along the analysed sequences. It defines three degrees of conservation that are each marked with a distinct symbol: asterisk (strong conservation), colon (medium conservation), and single dot (weak conservation). For simplicity, in ExpressInHost we only consider the highest degree of conservation of amino acids, which are the positions marked by an asterisk.

In this tuning mode, `Process_funct.cpp` performs the following steps: i) it aligns the codon sequences (target and orthologous) following the amino acid sequence alignment obtained with Clustal; ii) it identifies the codon positions of highly conserved amino acids according to the Clustal alignment, and it tags those positions; iii) it reads each nucleotide sequence (target and orthologous) codon by codon. If the codon position is not tagged, it replaces the codon by the fastest synonymous one (rank index = 1) in the host tRNA table. Otherwise, it searches for the host synonymous codon of closest rank index to the native one. Therefore, the tagged positions are preserved from speed maximisation, and instead they are tuned in the same way as in tuning mode 1 (Direct mapping).

**Mode 3: Optimisation and conservation II.** This tuning mode requests the same set of input files as for tuning mode 2: a set of nucleotide sequences that contains the target mRNA and its orthologous sequences, as well as the corresponding Clustal amino acid alignment. In this tuning mode, however, the Clustal alignment is exclusively used to align the nucleotide sequences.

In this tuning mode, `Process_funct.cpp` performs the successive steps: i) it aligns the codon sequences (target and orthologous) following the amino acid sequence alignment obtained with Clustal; ii) it reads the aligned nucleotide sequences codon by codon, and at each codon position it determines whether a slow codon is consistently used across the entire set of mRNAs (i.e. if at least 75% of the codons aligned at this position are slow). If so, the position is tagged; iii) if the codon position is not tagged, it replaces the codon by the fastest synonymous one (rank index = 1) in the host tRNA table. Otherwise, it searches for the host synonymous codon of closest rank index to the native one. Therefore, the tagged positions are preserved from speed maximisation, and instead they are tuned in the same way as in tuning mode 1 (Direct mapping).

To enable comparative analysis of the whole set of orthologous sequences after tuning, and to take into account the fact that the position of the target sequence in the Clustal set depends on the alignment process, we choose to tune all input nucleotide sequences, instead of only tuning the target sequence. It means that the orthologous sequences provided in modes 2 and 3 for similarity analysis are also tuned for expression in the host. Therefore, the first output file produced by ExpressInHost contains the tuned nucleotide sequence for each nucleotide sequence provided (target sequence alone, or together with orthologous genes (relevant for the modes 2 and 3)). The second file indicates how different is each tuned sequence in comparison to its native version, by providing the percentage of codons that have not been changed in the tuning procedure. In total, up to 10 sequences can be tuned at the same time, which can for instance be 1 target sequence and 9 orthologous genes used for the similarity analysis.

Figure 2: Graphical user interface of ExpressInHost.

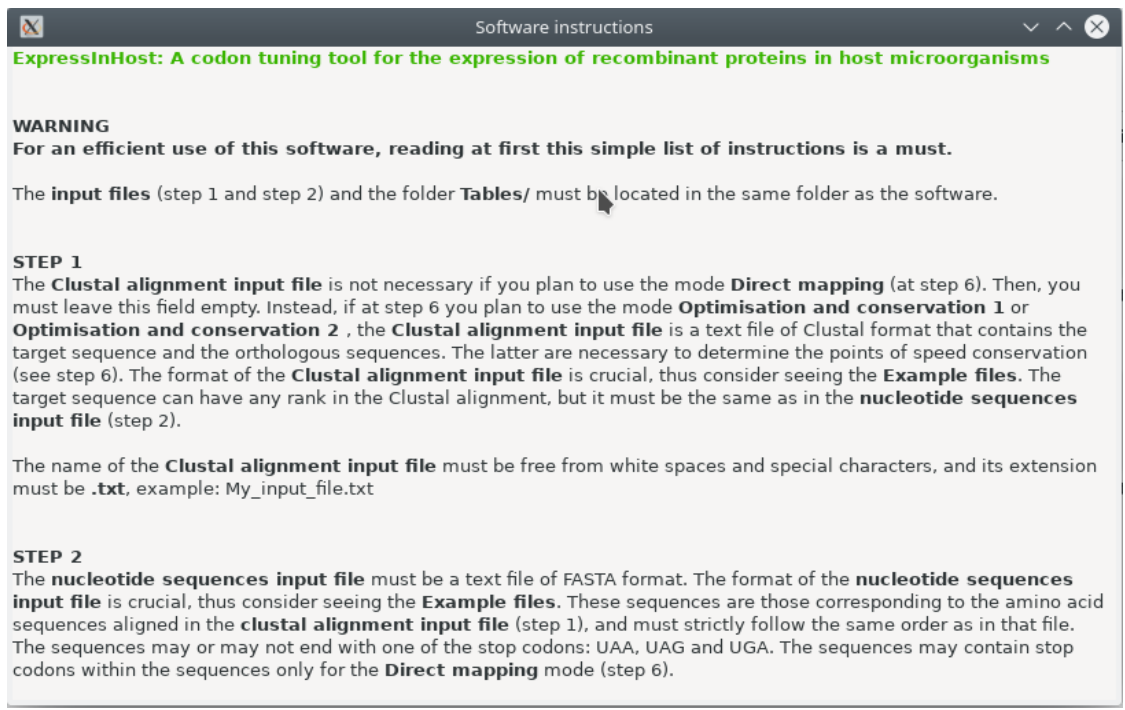

Figure 3: Expandable Instructions window of ExpressInHost.

```
Staphylococcus_aureus_recA          --------MDNDRQKALDTVIK----NMEKSFGKGAVMKLGDNIGRRVSTTSTGSVTLDN   48
Methanocaldococcus_jannaschii_DSM   ASIGELTEIDGISEKAAARIIEAARELCNLGFKSGTEVL---SQRKNIWKLSTGSKNLDE  120
Caenorhabditis_elegans_Rad51        TTRRELRNVKGISDQKAEKIMKEAMKFVQMGFTTGAEVH---VKRSQLVQIRTGSASLDR  165
Drosophila_melanogaster_Rad51       ATKKQLMAIPGLGGGKVEQIITEANKLVPLGFLSARTFY---QMRADVVQLSTGSKELDK  108
Saccharomyces_cerevisiae_Rad51      APRKDLLEIKGISEAKADKLLNEAARLVPMGFVTAADFH---MRRSELICLTTGSKNLDT  169
Arabidopsis_thaliana_Rad51          TPRKDLLQIKGISDAKVDKIVEAASKLVPLGFTSASQLH---AQRQEIIQITSGSRELDK  114
Danio_rerio_Rad51                   APKKELLNIKGISEAKADKILTEAAKMVPMGFTTATEFH---QRRAEIIQISTGSKELDK  112
Xenopus_laevis_Rad51                APKKELLNIKGISEAKAEKILAEAAKLVPMGFTTATEFH---QRRSEIIQISTGSKELDK  108
Homo_sapiens_Rad51                  APKKELINIKGISEAKADKILAEAAKLVPMGFTTATEFH---QRRSEIIQITTGSKELDK  111
Gallus_gallus_Rad51                 APKKELLNIKGISEAKADKILAEAAKLVPMGFTTATEFH---QRRSEIIQITTGSKELDK  111
                                            :  .        ::             .* ..  .         :   :**  **
```

Figure 4: Example of Clustal alignment.

## Quality control

The GUI is accompanied by an "Instructions window" (see Figure 3), and, in case a bug is detected, a "Debugging window" pops up and informs the user on the bug detected, and on the actions to be taken to sort the problem. This is enabled by a total of 25 tests throughout the software. They verify that the user has properly filled the input information in the graphical interface, and that the format of the input files provided fulfils the requirements. Just to give a few examples, the software verifies that the input files can be found in the directory, that they are not empty or contain unexpected characters in the sequences, and it checks whether the number of amino acid sequences and nucleotide sequences in the respective input files are the same.

**Case example**

ExpressInHost is released with an example (a set of input files) that can be tuned using any of the three tuning modes and of the four host microorganisms included in the software. Fully detailed instructions on how to run this example are given in the `Readme.md` file deposited together with the software. This example is based on the protein Rad51-RecA-RadA from *Homo sapiens*.

Here, to provide some further details on this example, we select a portion of 10 codons of the protein (codons 112 to 121), and we tune them for expression in the host organism *Escherichia coli*, using tuning mode 2 (Optimisation and conservation I). For this tuning mode, the software requests mRNAs orthologous to the target from different organisms. We use *Gallus gallus*, *Xenopus laevis*, *Danio rerio*, *Arabidopsis thaliana*, *Drosophila melanogaster*, *Saccharomyces cerevisiae*, *Caenorhabditis elegans*, *Methanocaldococcus jannaschii*, and *Staphylococcus aureus*. In Table 1, we compare the native sequence in *Homo sapiens* with the sequence output by the tuning mode 2, and with the sequence in which each codon has been substituted by the fastest synonymous one in the host (fully optimised). Full optimisation is not a mode implemented in our software, since our approach focusses on capturing key positions that may aid co-translational folding. However, it is useful to present it here for illustration purposes. During full optimisation, codons of rank index = 1 are systematically selected along the entire sequence, i.e. the speed is maximised throughout. In Table 1, we highlight with an asterisk the codons that have been tagged by the tuning mode 2, and we give their rank index. For those codons, translation speed has been matched to the native translation speed. In that specific example, we observe three different scenarios: i) codons are not tagged, and therefore they are the same to those resulting from full optimisation (codons 113, and 116-121); ii) codons are tagged and therefore their speed is not optimised (codons 114 and 115); iii) codons are tagged but their rank index is equal to one. Therefore, when mapping their translation speed into the host, the tuned codon is the same to that obtained through full optimisation (codon 112).

|  | Input: *Homo sapiens* | Output: *Escherichia coli* | |
| --- | --- | --- | --- |
| Position | Native/Rank index | Full optimisation | Optimisation and conservation I/Rank index |
| 112 | *CUU/1 | CUG | *CUG/1 |
| 113 | CAA | CAA or CAG | CAA or CAG |
| 114 | *GGU/0.46 | GGC | *GGU/0.53 |
| 115 | *GGA/0.44 | GGC | *GGU/0.53 |
| 116 | AUU | AUC | AUC |
| 117 | GAG | GAA | GAA |

| 118 | ACU | ACC or ACG | ACC or ACG |
|-----|-----|------------|------------|
| 119 | GGA | GGC | GGC |
| 120 | UCU | UCC | UCC |
| 121 | AUC | AUC | AUC |

Table 1: Codons 112 to 121 of the protein Rad51-RecA-RadA from *Homo sapiens* in its native, fully optimised, and tuned by mode 2 versions. The host organism used is *Escherichia coli*. Codons tagged in the mode 2 are highlighted by an asterisk and their rank index is indicated.

**Important effects**
The user should be aware of certain effects before running the software.

i) There are cases of synonymous codons decoded by different tRNAs, but with those different tRNAs being present at the same abundance (their GCN are the same). Hence, those codons have the same rank index, provided that their situation regarding base-pair wobbling is the same. We call such codons "equivalent codons". Additionally, for mode 1, and tagged codons in modes 2 and 3, in the (unlikely) case that in the host tRNA table synonymous codons of different rank index are equally distant to the rank index of the native codon, these host codons are also considered as "equivalent codons". For any of the tuning modes, when searching the appropriate tuned codon, the software randomly picks among "equivalent codons", to avoid codon usage bias. This has three consequences. First, when the software directly maps the translation rank index of a native codon into the host (mode 1, and tagged codons in modes 2 and 3), if more than one codon can be chosen in the host ("equivalent codons"), and if the native codon is one of these options, it will not preferably be selected. Second, when repeatedly tuning a nucleotide sequence with the same tuning mode, and the same native and host organisms, the successive outcomes are most likely different. Third, if the native and the host are the same organism, the output sequence will most likely be different from the input one upon calling mode 1.

ii) For the modes 2 and 3, the number of orthologous sequences used for the similarity analysis has an effect on the outcome. The more sequences selected (up to ten including the target, in ExpressInHost), the better the statistics to assess the conservation, i.e. conservation of amino acids in mode 2, and conservation of slow translation codons (low speed indices) in mode 3. Also, the phylogenetic diversity of the native organisms in the set of orthologous sequences directly impacts on the number of tagged codons. These two consequences are illustrated in Figure 5, for the tuning mode 2. The figure is based on the case example provided with the software, and mentioned above (Rad51-RecA-RadA protein from *Homo Sapiens*). In Figure 5, starting from a set of 10 orthologous sequences for the Rad51-RecA-RadA protein, the set is progressively reduced by removing one by one the orthologous proteins. At each step, the organism whose protein is removed is the most phylogenetically distant one, in comparison to *Homo sapiens*. By analysing the amino acid conservation at each step, we clearly see that picking a set of 10 sequences for highly diverse proteins (leftmost point) leads to much lower amino acid conservation than picking a set of 2 closely related species (rightmost point). As a consequence, respectively either 10% or 100% of the codons will be tagged, meaning that the tuned translation speed profile for these two extreme scenarios will be considerably different.

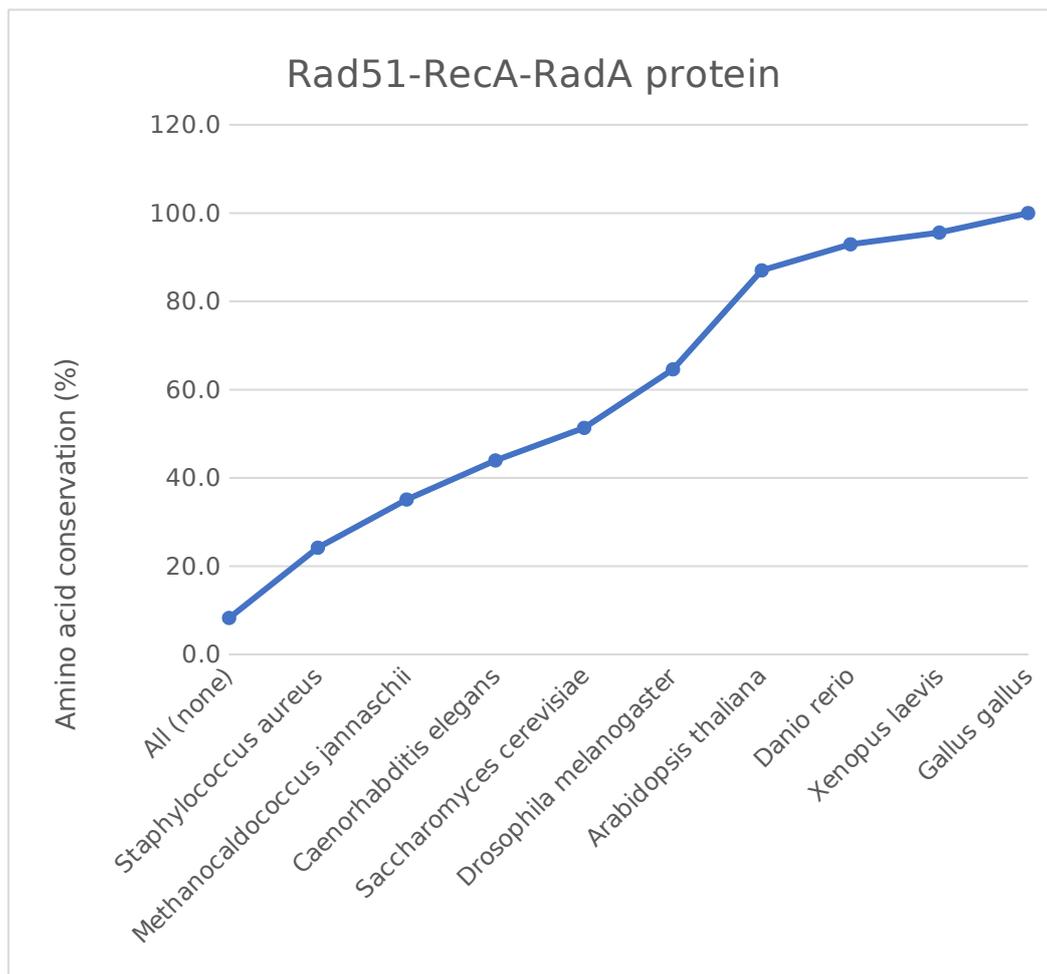

Figure 5: Effects of the number of proteins and phylogenetic diversity in the input set of orthologous genes, shown for mode 2. The name of the native organism whose protein is removed from the set of orthologous sequences is shown on the abscissa, while the amino acid conservation is measured on the ordinate. The less proteins considered and the closer they are related, the more amino acids are conserved.

# (2) Availability

The software release can be downloaded from the gitlab repository https://gitlab.com/a.raguin/expressinhost
Detailed instructions for installation and running are provided in the `Readme.md` file of the project.

**Operating system**
*nix (tested on Ubuntu 20.04 LTS)
Windows (tested on Windows 10)

**Programming language**
C++ making use of the GTK3 library for the user interface.

**Additional system requirements**
No

**Dependencies**
gtkmm 3.0
Windows only (Visual C++ redistributable for Visual Studio 2019)

**List of contributors**
Raguin, Adélaïde: Design, implementation, commenting, testing. Department of Computer Sciences, Institute of Computational Cell Biology, Heinrich-Heine University, 40225 Düsseldorf, Germany

**Software location:**
  **Archive** (e.g. institutional repository, general repository) (required – please see instructions on journal website for depositing archive copy of software in a suitable repository)
    **Name:** Zenodo
    **Persistent identifier:** 10.5281/zenodo.5113427
    **Licence:** Creative Commons Attribution 4.0 International
    **Publisher:** Adélaïde Raguin
    **Version published:** v0.1
    **Date published:** 19/07/2021
  **Code repository** (e.g. SourceForge, GitHub etc.) (required)
    **Name:** Gitlab
    **Identifier:** https://gitlab.com/a.raguin/expressinhost
    **Licence:** GNU version 3
    **Date published:** 19/07/2021
  **Emulation environment** (if appropriate)
    **Name:** The name of the emulation environment

***Identifier:*** The identifier (or URI) used by the emulator
***Licence:*** Open license under which the software is licensed here
***Date published:*** dd/mm/yy

**Language**
English

## (3) Reuse potential

ExpressInHost is conceived as a tool easily and freely accessible for users from any research or industrial domain interested in recombinant protein expression in a host microorganism. The GUI has been implemented precisely to facilitate the access to the software.

Two levels of further development are envisaged. First, from the GUI (in step 4) users can call new input table of tRNA data, in case their native organism of interest is not listed in the dropdown menu in step 3. However, the software does not include a GUI access to call host organisms different from those proposed. We made that choice since less diversity exists for the host microorganisms and the software already offers four of the most popular ones (*E. coli*, *S. cerevisia*, *K. pastoris*, and *B. subtilis*). Second, the source codes are extensively commented, such that users with programming background in C++ can easily get started with further development of the software. For instance, they might wish to process sets of orthologous genes larger than 10, or to express proteins in host microorganisms distinct to those currently proposed.

For support and minor extensions of the code, A.R. is available by email at the corresponding address. A.R. is also available by email if users encounter unexpected issues while using the software.

## Acknowledgements

### Funding statement
This work was performed as part of the Innovate UK project "Predictive optimisation of biocatalyst production for high-value chemical manufacturing" (Project Number TP101439). The current position of A.R. is funded by the German federal and state programme Professorinnenprogramms III for female scientists.

### Competing interests
The authors declare that they have no competing interests.

---